# Agent Choice via Quantum Flux in Living Systems

Ruth E. Kastner
University of Maryland, College Park
January 13, 2026

*ABSTRACT.  A basic model is provided that places active, intentional choices by biological  organisms on a solid physical footing. The model is provisionally called "Agent Choice via Quantum Flux."  It brings to bear specific physics on living systems in a way that allows for intentional choices not pre-determined by physical laws, but remaining consistent with those laws. It does so by exploring a possible many-to-one relation of quantum states to agent choices, with a parallel to the relation of thermodynamic microstates to macrostates.*

It is often assumed that free will is not licensed by physics. In particular, if the physical laws are deterministic, then all events are predetermined, in a Laplacian manner, and thus there are no real "live choices": organisms are fated to whatever outcomes are dictated based on initial conditions and those deterministic laws. This would imply that at the most basic levels of life, organisms are essentially machines incapable of individual initiatives that would constitute active self-organization and other self-asserting and self-sustaining behaviors (let alone such higher-level pursuits as human creativity). in other words, such laws preclude genuine volition or agency.  On the other hand, if physics is actually indeterministic, as would seem to be the case for quantum theory, then those outcomes are still subject to the Born rule for probabilities of their occurrence. Does this really mean however, that biological systems (or any physical system for that matter) are "slaves to the Born rule" as Sider (2005) asserts? In other words, would it be correct to conclude that what we call "living matter" can have no genuine claim to authentic agency--i.e., the capability to initiate actions of its own choosing that are neither dictated nor made chaotically "random" by physical law?

Our proposal counters the above pessimistic conclusion and offers a coherent physical basis for authentic agency on the part of living systems. It does so by making use of an important loophole first identified in Kastner (2016), which considers the "room for volition" provided by the indeterminism inherent in the Born rule's probabilistic description, yet refutes the usual assumption that all entities must be "slaves to the Born rule".[1] Specifically, it is not necessarily the case that the choices of a complex biological system, such as a cell, are represented by observables on Hilbert space applying to individual quantum systems in specific stable states. Consider, for example, a bacterium faced with a choice whether to move to the right or to the left. Contrary to the tacit assumption underlying claims that organisms are "slaves to the Born rule", it is not obligatory (nor probably even correct) to model such a choice as a quantum observable with outcomes corresponding to eigenstates "R" or "L" applying to the bacterium. This is because a meso- or macro-size organism like a bacterium makes that choice over a period of time during which many of its component quantum systems and states are undergoing rapid change. Specifically, nutrients are circulating and cellular subsystems are undergoing continual change, such that no single quantum system in a particular state (such as a molecular contituent of the intracellular fluid) is uniquely instantiating the "choice" of the organism. Instead, the organism's "choice" state is being supporting by a myriad of individual quantum systems and states that are changing over a relatively long time period, by quantum standards. Thus, the key point is that there is likely a many-to-one relation of quantum observables and states to a macroscopic "choice state" such as R/L. In a nutshell, a given choice does not correspond to a quantum observable but rather to a *family* of quantum observables and associated states, any of which could instantiate the choice.

---

[1] Here, we eschew hidden variables theories, which produce a deterministic account of outcomes that leave no room for agent choices.

There is an apt analogy here between a thermodynamic macrostate and its component microstates: specifically, the organism's choice state--being "ready to choose" among clearly defined alternatives--is the macrostate, and the various quantum observables and quantum states instantiating it are the supporting microstates. These microstates are in constant flux, much as a gas in a volume V at a given temperature is transitioning rapidly through different microstates while remaining in the same macrostate. Further, in correspondence to an actual implemented choice, transitions of the gas to a different macrostate are also supported by many different microstates. Based on an analogous "quantum flux" of the contitutents of a biological organism, the present proposal develops a model in which individual component quantum systems remain subject to the Born rule for any given measurement of a particular observable, but that corresponds only to a "microstate" in the above sense, so does not constrain the full scope of possibilities for instantiation of the choice "macrostate." Thus, meso- or macro-scale biological systems can evade the strictures of the Born rule to a large enough degree to essentially remove it as a constraint.

The model, which we call "Agent Choice via Quantum Flux" (ACQF), makes use of the idea that a meso – or macroscopic system, such as a bacterium, may have stable choice states that are instantiated by many individual quantum states that change over a relatively extended period of time during which a given choice remains available, and where choice outcomes are also instantiated by many different states, observables, and systems. Even if the actual choice can be attributed to an outcome of a measurement on an individual quantum system (e.g. molecule), that need not lead to a violation of the Born rule, since the same state and outcome need no longer be in play for the same macro-level choice to be made again. In other words, the next time the organism is faced with the same choice, it may be instantiated by different quantum states and different observables; so that even if the same choice is made over and over again, it does not violate the Born rule--since that applies only to a particular state and observable.

A specific example may help to illustrate this point. Consider a single-cell bacterium with a food supply to its right. It happens to encounter nutrients coming from that side, and this "primes" the bacterium to prefer the rightward direction. In a deterministic mechanical account, this process of chemotaxis would be modeled as a passive response on the part of the bacterium to laws and initial conditions, such that it has no actual choice as to its future direction of motion. However, we consider a different view here, and interpret chemotaxis as having elements of volition and intent based on information received (nutrients coming from the right). This includes the availability of "live options" as to which direction to choose; the bacterium is not mandated by physical law to move to the right. The single macro-choice presented, i.e., choose "right or left"--call it R/L--is instantiated by many possible sets of quantum states $|\Psi_{ij}\rangle$ and observables $O_{ijk}$. Here, *i* indexes the specific quantum system (e.g., molecule) in the state, *j* indexes the specific state acted on by the observable, and *k* indexes differing observables (such as angular momentum). We also need to represent the chosen outcomes by a set of states $|\Phi_{ikm}\rangle$. where m indexes the particular outcome state chosen. The ranges of the indices are in general very large, i.e. the number of component quantum systems N>>1 and the number of supporting states and observables similarly large. Thus we have a many -to-one relation of $\{|\Psi_{ij}\rangle, O_{ijk}, |\Phi_{ikm}\rangle\}$ to the macro-choice R/L.

For illustration, a "toy model" could involve just three observables, three different ready states and outcome states, and three quantum systems such as simple molecules. The observables could be the 3 spin-operators $S_x$, $S_y$, $S_z$; the three "ready" states could be $|\alpha+\rangle, |\beta+\rangle, |\gamma+\rangle$ where the Greek letters are mutually orthogonal directions (differing from x,y,z) and the plus signifies "up" along those directions. Outcome states implementing a decision "R" could be $|x+\rangle, |y+\rangle, |z+\rangle$, and for a decision "L" the opposite, i.e., $|x-\rangle, |y-\rangle, |z-\rangle$. Thus a ready state at the level of the bacterium can be instantiated by any of the three molecules in any of the three states above, and a choice can be made through the action of any of $S_x$, $S_y$, $S_z$ on any of the three molecules.

A nutrient gradient places any one of the three molecules (perhaps components of the cytoplasm) in one of the ready states above. The bacterium thus enters a macro-"ready" state supported by any of the individual ready states, as described above. This state can be maintained as long as any of the molecules is in any of the micro-ready states; they may undergo change among the different possible ready states as the organism "considers its options."  The choice can be implemented by the action of any of the three observables on any of the molecules in any of the ready states. So, for example, suppose the bacterium's internal systems implement a measurement of the observable $S_x$ on molecule 1 in ready state $|\alpha +\rangle$; the Born rule then applies to this single measurement. However, the outcome is indeterministic, such that no mechanical law determines the outcome. As discussed in Kastner (2016), this is what leaves room for volition on the part of the cell: by way of the molecule, it can, in theory, choose to actualize $|x +\rangle$ rather than $|x -\rangle$, and thus move to the right through an act of volition.[2]  Repeated choices to move to the right need not violate the Born rule, since the same macro-choice can be instantiated by different states and different observables. Of course, for the case of only 3 systems, states and observables, this process could lead to an observable Born rule violation. But actual cells have many orders of magnitude larger numbers of degrees of freedom, relevant observables, and ready states.

Thus, this model provides a way for intent and volition to be active in living systems, by recognizing that choices at the level of such macro-systems are not limited to single quantum observables and states subject to replicable cumulative deviations from the Born rule. Instead, the choices are supported by many distinct systems, states, and observables, analogous to the contrast between a thermodynamic macrostate and its supporting set of microstates. The model therefore provides a way to further entertain Freeman Dyson's intriguing suggestion that " … our consciousness is not just a passive epiphenomenon carried along by the chemical events in our brains, but is an active agent forcing the molecular complexes to make choices between one quantum state and another. In other words, mind is already inherent in every electron, and the processes of human consciousness differ only in degree but not in kind from the processes of choice between quantum states which we call "chance" when they are made by electrons." (Dyson 1979, 249)

Of course, many important questions remain: How is it that the cell senses the information represented by the choice-ready states of individual quantum systems? How exactly are specific quantum observables brought to bear to implement the choice, and their measurements carried out by the cell?  The current proposal does not directly address those questions, which remain for further study, but provides a starting point for a coherent account of intent and volition in living systems, which shows that they need not be viewed as either passive"cogs in a deterministic machine" nor as "slaves to the Born rule" in an indeterministic account.

---

[2] In fact, an argument can be made, based on Leibniz's Principle of Sufficient Reason, that the lack of a deterministic law dictating a particular outcome requires an act of volition in order for a particular outcome to occur!